\documentstyle[multicol,aps,prb,epsf]{revtex}

\begin{document}

\draft

\title{An electron correlation originated negative magnetoresistance 
in a system having a partly flat band
}

\author{
Ryotaro Arita, Kazuhiko Kuroki and Hideo Aoki
}
\address{Department of Physics, University of Tokyo, Hongo, Tokyo
113-0033, Japan}

\date{\today}

\maketitle

\begin{abstract}
Inspired from an experimentally examined organic conductor, 
a novel mechanism for negative magnetoresistance is proposed 
for repulsively interacting electrons on a lattice 
whose band dispersion contains a flat portion 
(a flat bottom below a dispersive part here).  
When the Fermi level lies in the flat part, the electron correlation 
should cause ferromagnetic spin fluctuations to develop with an 
enhanced susceptibility.  A relatively small magnetic field will 
then shift the majority-spin Fermi level to the dispersive part, 
resulting in a negative magnetoresistance.
We have actually confirmed the idea by calculating the conductivity 
in magnetic fields, with the fluctuation exchange approximation, 
for the repulsive Hubbard model on a square lattice having 
a large second nearest-neighbor hopping.  
\end{abstract}

\medskip

\pacs{PACS numbers: 71.10.Fd, 74.20.Mn}
\begin{multicols}{2}
\narrowtext

Negative magnetoresistance has provided fascination in 
diverse classes of systems, such as impurity band in semiconductors 
or Mn oxides.  The problem is the interplay of the transport 
and the spin structure, and how the spins and carriers respond to 
magnetic fields can vary from system to system.    
In the physics of correlated electron systems, spin is 
a key ingredient.  In fact, itinerant (metallic) ferromagnetism 
has been a central problem from the days of 
Kanamori\cite{Kanamori}, Hubbard\cite{Hubbard63} and 
Gutzwiller\cite{Gutzwiller}.
It is becoming increasingly clear that the criterion for 
the ferromagnetism becomes stringent as one incorporates correlation effects, 
but that a large density of states at the Fermi level does favor 
ferromagnetism.  

In this paper, we propose a 
novel mechanism for negative magnetoresistance, 
which is realized as a combined effect of 
the electron correlation (proximity to ferromagnetism) and the 
band structure (coexistence of flat and dispersive parts).  
The idea is the following: 
We consider repulsively interacting electrons 
on a band, whose one-electron dispersion
has a {\it flat} part in an otherwise dispersive band.  
For low electron densities 
with the Fermi level $E_F$ located near the flat part of the band, 
the system should have a strong tendency toward ferromagnetism.  
A large density of states $D(E_F)$ at the bottom of the band
is indeed a situation 
originally considered by Kanamori\cite{Kanamori} and received a renewed 
interest recently for two and three dimensional systems 
\cite{Hlubina1,Hlubina2,Arita-Onoda} as well as in one dimensional systems 
\cite{Daul-Noack,Arita-Kusakabe}.
When a magnetic field is applied, 
a weak field is then enough to drive the system 
into a significant magnetization (spin polarization), 
since the magnetic susceptibility is enhanced due to
ferromagnetic fluctuations.  
The $E_F$ for the majority spins will then be shifted into the
dispersive (i.e., lighter-mass) part of the band, 
and we expect that the system becomes more
conductive for relatively weak magnetic fields.
When the polarization is small enough there is a possibility that 
the effect of the increase in $v_F$ of the majority spin may be compensated 
by the decrease in $v_F$ of the minority spin, 
but such a compensation will not occur when the polarization is sufficiently 
large due to the enhanced susceptibility, since not only $v_F$ but also the 
number of the majority spin increase in the presence of the magnetic field.  
Further, there will be fewer electron-electron scatterings for a polarized 
state, at least for the Hubbard model where only up and down 
spin electrons interact. These will make the system more conductive.
This idea has been inspired from a 
recent experimental result on a certain class organic 
conductors, called $\tau$-type conductors, 
for which Murata {\it et al}\cite{Murata} have observed 
negative mangetoresistance, see below.  
The band structure calculation indicates that 
the dispersion is indeed flat at the bottom 
along $k_x$ and $k_y$ directions.  

\begin{figure}
\epsfxsize=8cm 
\epsfbox{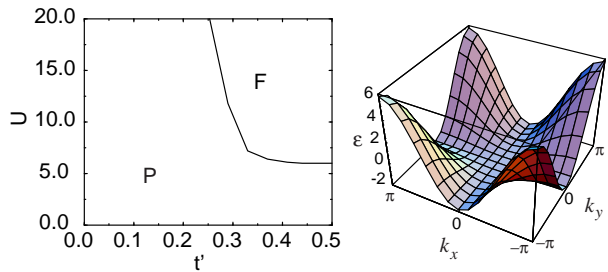}
\caption{The phase diagram of the $t$-$t'$ Hubbard model,
determined by the exact diagonalization of 
8 electrons in 4 $\times$ 4 system.  
The right panel is the one-electron dispersion for 
$t=-1, t'=0.5$.
}
\label{phase}
\end{figure}

In the following, we confirm this idea 
for the single-band Hubbard model, a simplest model for repulsive electron
correlation, 
on a simplest flat-bottomed tight-binding model, 
the square lattice having a large second
nearest neighbor hopping, $t'$, along with $t$ between nearest neighbors.  
The one-electron dispersion becomes
\begin{equation}
\varepsilon_{{\bf k}}^0=
2t \left( \cos k_x + \cos k_y \right)
+4t' \cos k_x \cos k_y , \nonumber
\end{equation}
and the dispersion is flat along $k_x$ and $k_y$ 
(right panel of Fig. \ref{phase}) for $t'\simeq 0.5|t|$ 
with the van Hove singularity coming down to the bottom.  

For $t'\simeq 0.5$ (we take $|t|=1$ hereafter), 
Hlubina has shown, with the T-matrix 
approximation for the Hubbard model, that the ground state is fully 
spin-polarized for $n\sim 0.4$.\cite{Hlubina1,Hlubina2} 
Since the T-matrix approximation is valid only in the low 
enough densities and for small $U$, 
here we start with obtaining the diagram for large $U$
with the exact diagonalization of a finite system 
(4$\times$4 sites with 8 electrons with the band quarter filled).  

Detection of ferromagnetic states from total ${\bf S}^2$ in 
finite systems requires some care as we have revealed in previous 
publications\cite{Arita-Kusakabe}.  
Namely, ferromagnetic states in itinerant electron systems 
can accompany a spiral spin state in finite systems.  The spiral 
state is a spin singlet state having the spin correlation wave 
length as large as the 
linear dimension of the system, which can lie in energy below the 
ferromagnetic state.  
We found that the ground state of the 16-site system has 
always ${\bf S}^2=0$ in periodic boundary conditions in
both $x$ and $y$, but if we look at the spin structure 
$S({\bf q})$ 
(Fourier transform of $\langle {\bf S}_i \cdot {\bf S}_j \rangle$) 
the state turns out to be indeed spiral 
(i.e., $S({\bf q})$ peaked at $(0,\pm\pi/L)$ and $(\pm\pi/L,0)$) 
for large $U$ and $t'\simeq 0.5$.

Although we can identify the spiral state as ferromagnetic in the 
thermodynamic limit, it has been shown\cite{spiral2}
that a faster approach to the thermodynamic limit is attained 
if we adopt an appropriate boundary condition 
(periodic in one direction and antiperiodic in another) 
to selectively push the spiral state above the ferromagnetic 
state in energy.  
If we look at the phase diagram thus obtained against $t'$ and $U$ 
in Fig. \ref{phase}, 
the fully polarized ferromagnetic region is seen to exist 
for $t'\sim 0.5$ and $U>6$. 

Keeping this phase diagram in mind, 
we now discuss the conductivity in magnetic fields $B$.  
The conductivity of the interacting system is calculated 
here by the fluctuation exchange approximation (FLEX).
The FLEX, introduced by Bickers {\it et al.}\cite{FLEX}, 
treats spin and charge fluctuations by starting from a set of 
skeleton diagrams for the Luttinger-Ward functional, 
based on the idea of Baym and Kadanoff\cite{Baym}.  
Then a ($k$-dependent) self energy is computed from 
RPA-type bubble and ladder diagrams self-consistently.  

The dc conductivity is given in the Kubo formula as
\begin{eqnarray*}
\sigma_{\mu\nu}=\lim_{\omega \rightarrow 0}
{e^2}\sum_{\sigma\sigma'}
\int \frac{d{\bf k}d{\bf k'}}{(2\pi)^6}
v^0_{{\bf k}\mu}v^0_{{\bf k'}\nu}
\frac{{\rm Im}K_{{\bf k}{\bf k'}\sigma\sigma'}(\omega+i\delta)}{\omega},
\end{eqnarray*}
where $K_{{\bf k}{\bf k'}\sigma\sigma'}(\omega+i\delta)$ 
is the Fourier component of the retarded two-particle Green's function 
and $v^0_{{\bf k}\mu}=\partial \varepsilon^0_{\bf k}/\partial {\bf k}_\mu$ 
is the unperturbed velocity.  We set $\hbar=1, k_B=1$ hereafter.
If we follow Eliashberg\cite{Eliashberg} for the 
analytic continuation of 
$K(\omega + i\delta)$ from $K(i \omega_n)$, where $\omega_n$ is 
the Matsubara frequency, the conductivity per spin reads, 
for long enough life time of the quasi-particle,
\begin{eqnarray*}
\sigma_{x x}={e^2}\int \frac{d{\bf k}}{(2\pi)^3}
\left( -\frac{\partial f}{\partial \varepsilon}
\right) \frac{v_{{\bf k}x}^*J_{{\bf k}x}^*}{2\gamma_{\bf k}} ,
\end{eqnarray*}
where $v_{{\bf k}x}^*$ is the dressed velocity, 
$J_{{\bf k}x}^*$ the current, 
$\gamma_{\bf k}$ the damping constant
of the quasiparticle, and $f$ the Fermi distribution function.  

Note that the conductivity of a clean system of interacting electrons 
can be finite even though the interaction is an internal force, 
since the momentum is dissipated through Umklapp processes. 
It has been shown by Yamada and Yosida\cite{Yamada} 
that the conductivity diverges, as it should, when the Umklapp
processes are turned off only if we consider 
the vertex correction for the current $J_{\bf k}$ appropriately, 
so the correction should be included for a consistent treatment.  
Within the FLEX, which is a conserving approximation,
there are three types of diagrams
for irreducible vertex\cite{KonKanUe}, namely two
Aslamasov-Larkin (AL) type diagrams and
one Maki-Thompson (MT) type one.
Kontani {\it et al}\cite{KonKanUe} have shown that, 
when antiferromagnetic fluctuations are dominant, 
the AL contribution can be neglected.  
We can extend this argument to show that the AL term is also negligible 
when ferromagnetic fluctuations are dominant.  
Hence we consider only the MT term.\cite{ALcomm}

Let us now discuss the conductivity in magnetic fields.  
We have then to add the Zeeman term, 
$
h\sum_{{\bf k}\sigma}{\rm sgn}(\sigma)
c_{{\bf k}\sigma}^\dagger c_{{\bf k}\sigma},
$
to the Hamiltonian, 
where $h\equiv g\mu_B B$ is the Zeeman energy with $g\simeq 2$.  
Green's function and other quantities then become $\sigma$-dependent.  
To concentrate on the effect of the Zeeman splitting we assume here 
that the direction of the magnetic field is parallel to the
current, so that we do not have to take account of the effect of the field 
on orbital motions. 

The conductivity is obtained from the 
Bethe-Salpeter equation, which 
is a simple extension of the equations derived by
Kontani {\it et al} for the spin-independent case\cite{KonKanUe} 
to the present spin-dependent case.  
We end up with the diagonal conductivity, 
\begin{eqnarray}
\sigma_{xx}&=&{e^2}
\sum_{{\bf k},\sigma} \int \frac{d \varepsilon}{\pi N}
\left(-\frac{\partial f}{\partial \epsilon}\right)
\left\{
|G_{{\bf k}\sigma}(\varepsilon)|^2 
v_{{\bf k}x\sigma} J_{{\bf k}x\sigma}(\varepsilon) \nonumber \right.\\ 
&&\left. -{\rm Re} \left[G_{{\bf k}\sigma}^2(\varepsilon)v_{{\bf k}\sigma}^2
(\varepsilon)\right] \right\} , \label{sigma}\\
&J_{{\bf k}x\sigma}&(\omega)=v_{{\bf k}x\sigma}(\omega) \nonumber\\
&&+ \sum_{{\bf q}\sigma'}\int \frac{d\varepsilon}{2\pi N} 
\left[ {\rm cotanh}\frac{\varepsilon-\omega}{2T}
-\tanh\frac{\varepsilon}{2T}\right] \nonumber\\
&&\times {\rm Im}V_{{\bf k-q},\sigma\sigma'}
(\varepsilon-\omega+i\delta)|G_{{\bf q}\sigma'}(\varepsilon)|^2
J_{{\bf q}x\sigma'}(\varepsilon)\nonumber,
\end{eqnarray}
where $N$ is the number of sites, $T$ the temperature, 
$G_{{\bf k}\sigma}(\omega)$ the dressed Green's function, and the velocity 
$v_{{\bf k}x\sigma} = (\partial/\partial k_x) [\varepsilon_{\bf k}^0
+{\rm Re}\Sigma_{{\bf k}\sigma}(\omega=0)]$ 
with $\Sigma_{{\bf k}\sigma}(\omega)$ being the self energy. 

The kernel, $V_{{\bf k}\sigma\sigma'}(\omega)$, 
which contains the effect of fluctuation exchanges, 
can be obtained by an analytic continuation of
$V_{{\bf k}\sigma\sigma'}(i\omega_n)$\cite{Vidberg}, 
\begin{eqnarray*}
V_{{\bf k}\uparrow\uparrow}(i\omega_n)&=&
\frac{U^2\chi^0_{{\bf k}\uparrow\uparrow}(i\omega_n)}
{1-U^2\chi^0_{{\bf k}\uparrow\uparrow}(i\omega_n)
\chi^0_{{\bf k}\downarrow\downarrow}(i\omega_n)}\\
&&-\frac{U^2}{2}\chi^0_{{\bf k}\uparrow\uparrow}(i\omega_n),\\
V_{{\bf k}\uparrow\downarrow}(i\omega_n)&=&
\frac{U^2\chi^0_{{\bf k}\uparrow\downarrow}(i\omega_n)}
{1-U\chi^0_{{\bf k}\uparrow\downarrow}(i\omega_n)}
-\frac{U^2}{2}\chi^0_{{\bf k}\uparrow\downarrow}(i\omega_n)+U,\\
\chi^0_{{\bf k}\sigma\sigma'}(i\omega_n)&=&-\frac{T}{N}
\sum_{k}G_\sigma(k+q)G_{\sigma'}(k),\\
\end{eqnarray*}
where $k\equiv ({\bf k}, i\omega_n)$ in the last line.

\begin{figure}
\epsfxsize=5cm 
\epsfbox{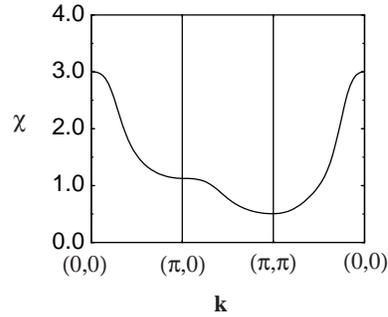}
\caption{
The wave number dependence of $\chi^{\rm RPA}_{\bf k}$ for
$t'=0.5, U=2, n=0.4, T=0.1$.
}
\label{chi-kT01}
\end{figure}

Let us now present the results.  
We take the case of $t'=0.5, U=2$ with the band filling $n=0.4$, 
which falls upon the paramagnetic region close to the ferromagnetic 
boundary in the phase diagram, Fig.\ref{phase}.  
We first check that we do have strong ferromagnetic fluctuations.  
Figure \ref{chi-kT01} shows the wave number dependence 
of the spin susceptibility, 
$\chi^{\rm RPA}_{{\bf k}} = 
2\chi^0_{{\bf k}}(\omega=0)/(1-U\chi^0_{{\bf k}}(\omega=0))$ 
for $T=0.1$. We can see that there is indeed a peak around $\Gamma$ 
(${\bf k}={\bf 0}$).

\begin{figure}
\epsfxsize=5cm 
\epsfbox{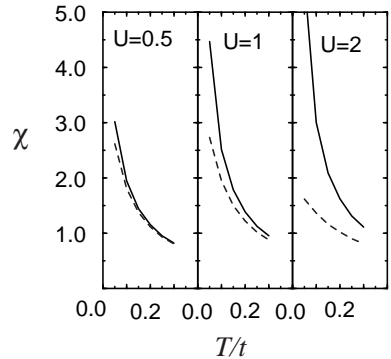}
\caption{
The temperature dependence of 
the static magnetic susceptibility $\chi^{\rm FLEX}$(dashed line)
and $\chi^{\rm RPA}_{{\bf k}={\bf 0}}$(full line)
for $U=0.5, 1.0, 2.0$ with $t'=0.5, n=0.4$.
}
\label{chi}
\end{figure}

Let us next focus on the static, uniform magnetic susceptibility.  
Since we are sitting close to the ferromagnetic boundary, 
the susceptibility should be finite but enhanced.  
Here we compute the quantity in two ways: one is 
to calculate the derivative 
$\chi\equiv\partial \langle n_\sigma-n_{-\sigma}\rangle/\partial h$ 
with a finite difference $\delta h=0.005$, 
and the other is $\chi^{\rm RPA}_{\bf 0}$. 
If we look at their temperature dependence 
in Fig. \ref{chi} for $t'=0.5, n=0.4$ with 
various $0.5 \leq U \leq 2.0$, 
they both sharply increase for $T\rightarrow 0$. 
There is a deviation between $\chi$ and $\chi^{\rm RPA}_{\bf 0}$ 
for larger $U$.  The deviation itself indicates that the irreducible
four-point vertex  $\Gamma=\delta^2 \Phi/\delta G\delta G$ 
cannot be approximated by $U$.  
Since we are dealing with a case where 
the ferromagnetic phase appears for larger $U(>6\sim7)$, 
$\chi$ underestimates the effect of external magnetic field for large $U$.  

\begin{figure}
\epsfxsize=7cm 
\epsfbox{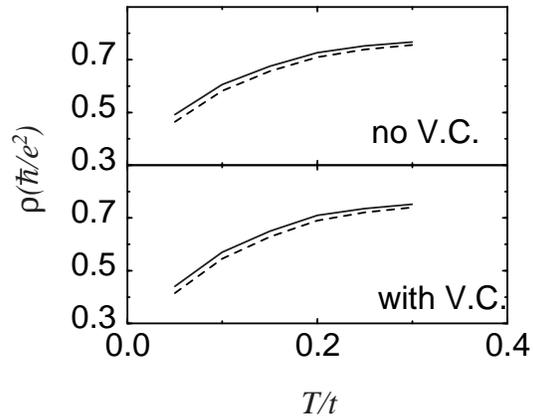}
\caption{
The temperature dependence of the resistivity of the $t$-$t'$ Hubbard model
with (bottom) or without (top) vertex corrections 
for $t'=0.5, U=2, n=0.4$ 
in zero magnetic field ($h=0$, full line) and 
in a magnetic field ($h=0.05$, dashed line).
}
\label{rho}
\end{figure}

Now, we come to the key result for the magnetoresistance, 
given in Fig. \ref{rho} for $t'=0.5$, $U=2$, $n=0.4$.  
The figure compares the resistivity against $T$ in the absence 
and in the presence of the magnetic field.  
We take the system size $N=64^2$, with $512$ 
Matsubara frequencies, which is checked to be sufficient 
for the temperature region studied here.  
We can see that we do have a negative magnetoresistance. 
The change in the resistance is of the 
order of 10\% for the Zeeman energy $h=0.05$.  
If we were not sitting close to the ferromagnetism, much 
larger fields would be required.  
The vertex correction does not alter the 
result significantly, which implies that Boltzmann's transport picture 
(with no vertex correction) already has the negative magnetoresistance.  
The negative magnetoresistance should become more 
prominent at lower temperatures where 
the ferromagnetic fluctuation increases.  
This is not too noticeable in the result, 
which should be an artifact: as mentioned above, 
the spin polarization is underestimated in the FLEX at low 
temperatures for $U=2$.

As touched upon at the beginning, 
Papavassiliou{\it et al.}\cite{Papa} 
has found experimentally that organic salts D$_2$A$_1$A$_y$, 
based on D (= P-$S$, $S$-DMEDT-TTF or EDO-$S$,$S$-DMEDT-TTF)
in combination with linear anions A (=AuBr$_2$, I$_3$, or IBr$_2$) 
with the fractional value of $y$ controlling the density of carriers 
are two-dimensional metals in the $\tau$ crystal form.  
The band structure of a single layer
of the $\tau$-phase, calculated with the
extended H{\" u}ckel method, contains a flat-bottomed band\cite{Papa2}.

\begin{figure}
\epsfxsize=3cm 
\epsfbox{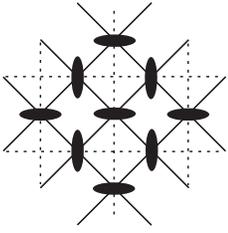}
\caption{
The in-plane molecular configuration in $\tau$-D$_2$A$_1$A$_y$.
The solid lines denote the nearest neighbor hopping while
dashed lines the second nearest neighbor hopping 
between face-to-face molecules.
}
\label{tau}
\end{figure}

In terms of the tight-binding model, 
we can regard that the in-plane molecular configuration is such that 
the next-nearest intermolecular 
hopping, $t_2 \simeq 0.02$ eV, 
appears in every other plaquetts in a checker-board manner 
on top of the nearest one $t_1\simeq 0.2$ eV 
(Fig. \ref{tau})\cite{Ducasse}.  
The checker-board makes the Brillouin zone 
folded, where the dispersion of the upper band in which $E_F$ resides is
like the square root of Eqn.(1) with $t' = 0.5$ when the 
splitting due to $t_2$ is small.  
This way we can have an anisotropically flat dispersion along $k_x, k_y$ 
with a large $D(E_F)$ at the bottom of the band, so 
a ferromagnetic component in the spin 
fluctuation exists for this dispersion, 
as we have checked with FLEX.  
If we naively plug in 1/8 
($\sim 0.05$ eV) 
of the width of the upper band 
calculated in ref. \onlinecite{Ducasse} for $t$,
the $h=0.05$ for $U=2$ in Fig. \ref{rho} corresponds 
to $B \simeq 30$ T. 
Since the divergence of $D(E_F)$ is weaker than that 
for Eqn.(1) because of the square root, 
large values of $U$ should be necessary to obtain
appreciable negative magnetoresistance.  
In organic materials $U$ may be indeed large, but we have
then to check whether the band mixing across the gap for
large $U$ can smear out the ferromagnetism.
Our preliminary result for the two-band model with the 
multi-band FLEX shows that the ferromagnetic component 
in the spin fluctuation 
does remain for $U \sim 4$, although it becomes weaker.  
Thus quantitative estimates of the magnetoresitance 
for the material should include these effects.

Murata {\it et al}\cite{Murata} have in fact 
observed a negative magnetoresistance 
for $B\sim$ several tesla with some hysteretic behaviors 
in this material\cite{Kurokicap}.  
Theoretically, however, we can argue that dominant
ferromagnetic fluctuations are not a necessary condition for 
the negative magnetoresistance conceived here.  
In the present scheme, the 
the system becomes more conductive in moderate $B$ due to a combination of 
a correlation-enhanced spin susceptibility 
and the flat-bottomed dispersion.  
So all we have to have about the magnetism is a large enough susceptibility.  
Thus the negative magnetoresistance observed by 
Murata {\it et al}\cite{Murata} is understandable provided 
the ferromagnetic {\it component} is present, if not dominant, 
in the spin fluctuation for the present mechanism to be relevant.

Although we have exemplified our idea so far in 
two dimensions, we believe that 
the present mechanism should be general, and can be found 
in other models with strong ferromagnetic spin fluctuations 
such as face centered cubic lattice, 
whose band structure has flat and dispersive parts as well.

We wish to thank Keizo Murata for extensive 
discussions and for showing us experimental data on the 
$\tau$-conductors, sometimes prior to publication.  
We also benefited from communications with 
Laurent Ducasse on the H\"{u}ckel parameters, and 
Hiroshi Kontani for the FLEX method.
R.A. is supported by a JSPS Research Fellowship for
Young Scientists, while 
K.K. acknowledges a Grant-in-Aid for Scientific
Research from the Ministry of Education of Japan.
Numerical calculations were performed at the Supercomputer Center,
ISSP, University of Tokyo.

\end{multicols}

\begin{references}
\bibitem{ALcomm} The peak around the $\Gamma$ (ferromagnetic) 
point in $\chi$ in Fig. \protect\ref{chi-kT01}
is not too sharp for $U=2$, so that 
the cancellation of the AL terms may not be perfect. 
However, we can expect that 
the vertex correction itself, hence the AL terms, will be small 
for $U\simeq 2$.
\end{references}
\end{document}